\documentclass[prl,preprint,showpacs,preprintnumbers,amsmath,amssymb,floatfix]{revtex4}

\usepackage{amsthm}
\usepackage{graphicx}% Include figure files
\usepackage{dcolumn}% Align table columns on decimal point
\usepackage{bm}% bold math
\usepackage{verbatim}

\begin{document}

\bibliographystyle{plainnat}
\title{Exact solution to a nearly parallel vortex filament mean-field theory}
\author{Timothy D. Andersen}
\email{andert@alum.rpi.edu}
\affiliation{Hampton, VA 23666}
\date{\today}

\begin{abstract}
Nearly parallel vortex filaments are a generalization of point vortices and describe many phenomena under conservation of angular momentum including vortices forming in deep ocean convection, magnetically confined plasmas, and the solar atmosphere. While point vortices represent perfectly straight, parallel lines of circulation, nearly parallel vortex filaments have some curvature due to internal viscosity.  They interact logarithmically but have a kinetic self-energy as well.  In this letter, I present an exact solution of a system of these filaments under angular momentum conservation in a mean-field theory.  I show that the filaments have an infinite order phase transition not present in the point vortex model.
\end{abstract}

\maketitle

Nearly parallel vortex filaments are a generalization of point vortices and describe many phenomena under conservation of angular momentum including vortices forming in deep ocean convection \cite{Andersen:2008a}, magnetically confined plasmas \cite{Andersen:2007b, Andersen:2008b}, and the solar atmosphere \cite{Julien:1996}. While point vortices represent perfectly straight, parallel lines of circulation, nearly parallel vortex filaments have some curvature due to core structure (viscosity and velocity) \cite{Klein:1991, Klein:1995}.  They interact logarithmically but have a kinetic self-energy as well.  In this letter, I present an exact solution of a system of these filaments under angular momentum conservation in a mean-field theory.  I show that the filaments have an infinite order phase transition temperature below which vortex formation is unstable.

The energy functional for $N$ periodic, piecewise linear vortex filaments of period $1$ with $M$ segments (each length $\delta=1/M$) is,
\begin{equation}
H(N) = \frac{1}{2}\sum_{i=1}^N\sum_{k=1}^M \left\{\alpha\frac{|z_i(k) - z_i(k+1)|^2}{\delta} +  p\delta|z_i(k)|^2 - \delta\frac{1}{2}\sum_{j=i+1}^N \log|z_i(k) - z_j(k)|^2\right\},
\end{equation} where $p$ is related to the angular velocity, $1/\alpha$ is the filament elasticity, and $z_j(k)=x_j(k) + iy_j(k)$ is a complex number indicating the position of the $k$th bead (link joint) on the $j$th filament in the xy-plane \cite{Lions:2000}.

The canonical ensemble is given by,
\begin{equation}
\Omega(N) = \int \mathrm{d}z_1\cdots\int \mathrm{d}z_N\, e^{- \beta H(N)},
\end{equation}  and the mean free energy $F(\beta) = -\beta^{-1}\log \Omega$.

The mean-field theory $|z_i(k) - z_j(k)|=|z_i(k) - \langle z_j(k)\rangle| = |z_i(k)|$ decouples the integrals.  Let $N=\Lambda$, the total circulation, be fixed, and let the number of filaments be large.  The mean-field partition function,
\begin{equation}
Z = \int \mathrm{d}z\, e^{-\frac{1}{2}\beta\sum_{k=1}^M \left\{\alpha\frac{|z(k) - z(k+1)|^2}{\delta} +  p\delta|z(k)|^2 - \delta\frac{\Lambda}{2} \log|z(k)|^2\right\}},
\end{equation} is exactly solvable.

To evaluate it, rewrite $Z$ as,
\begin{equation}
Z = \int \mathrm{d}z\, e^{-\frac{1}{2}\beta\sum_{k=1}^M \alpha/\delta|z(k) - z(k+1)|^2 +  p\delta|z(k)|^2}\prod_{k=1}^M  |z(k)|^{\delta\frac{\Lambda}{2}},
\end{equation} and consider the harmonic integral,
\begin{equation}
\Psi(J_1,\dots,J_M) = \int \mathrm{d}z\, e^{-\frac{1}{2}\sum_{k=1}^M \left\{\alpha\beta/\delta|z(k) - z(k+1)|^2 +  (\mu\delta - J_k)|z(k)|^2\right\}},
\end{equation} where $\mu=p\beta$, and evaluate $\Psi$ as a Gaussian integral,
\begin{equation}
\Psi = (2\pi)^{M}\prod_{k=1}^M \frac{1}{\mu\delta - J_k + K\lambda_k},
\end{equation} where $K=2\alpha\beta/\delta$ and $\lambda_k = 1 - \cos(2\pi(k-1)\delta)$ \cite{Berlin:1952}.  Through fractional differentiation by $J_k$, we evaluate $Z$,
\begin{equation}
Z = \prod_{k=1}^M \frac{\partial^{\delta\beta \Lambda/2}}{(\partial J_k)^{\delta\beta \Lambda/2}} \Psi(J_1,\dots,J_M).
\end{equation}  While fractional derivatives are often difficult to evaluate, this is a simple reciprocal and, letting $J_k=0$, we have
\begin{equation}
Z = \prod_{k=1}^M \frac{1}{(\mu\delta + K\lambda_k)^{\delta\beta \Lambda/2}}\Psi.
\end{equation}  The result is rigorous through analytic continuation.

Defining the free energy $F=-\beta^{-1}\log Z$, we have (dropping the harmonic factor $\Psi$),
\begin{equation}
F = \sum_{k=1}^M \delta \Lambda/2\log(\mu\delta + K\lambda_k),
\label{eqn:brokenF}
\end{equation} which I shall now evaluate in the continuum limit.

After an integration by parts, the mean field energy functional as $M\rightarrow\infty$ is,
\begin{equation}
H[\psi] = \frac{1}{2}\int_0^1 \mathrm{d}\tau\, \psi^\dagger(\mu - \alpha\beta\partial^2)\psi - \Lambda/2\log\psi^\dagger\psi,
\end{equation} where the vortex filament, $\psi(\tau)$, is a periodic function in $\mathcal{L}^2_{[0,1)}$.  (Under a Wick rotation, $H[\psi]$ is identical to the mean-field action of a quantum Coulomb gas of mass $\sqrt{p/\alpha}$ bosons in (0+1)-D spacetime \cite{Zee:2003}.) Therefore, in the free energy, the eigenvalues $\mu + K/\delta\lambda_k$ become those of the $\mu - \alpha\beta\partial^2$ operator.  Let $\omega^2=(2\pi)^2\alpha\beta$ and $m^2=\mu$.  Then the eigenvalues are $m^2 + \omega^2k^2$ (where $k$ is now a continuous momentum) and $F$ becomes,
\begin{equation}
F = \frac{\Lambda}{2}\int_0^1 \mathrm{d}k\, \log(m^2 + \omega^2k^2).
\end{equation}  

Evaluating the integral gives,
\begin{equation}
F = \frac{\Lambda}{2}\left[-2 + \log(m^2 + \omega^2) + 2\frac{m}{\omega}\tan^{-1}(\omega/m)\right].
\end{equation} The free energy shows a phase transition of infinite order (of Kosterlitz-Thouless \cite{Kosterlitz:1973}) and that the vortices are stable when $m/\omega$, $m$, and $\omega$ are all small.  If temperature is $T=1/\beta$ and we scale $\bar{\alpha}=(2\pi)^2\alpha$, the transition temperature is,
\begin{equation}
T_c = (p+\bar{\alpha})e^{2\sqrt{p/\bar{\alpha}}\tan^{-1}(\sqrt{\bar{\alpha}/p}) - 2}.
\end{equation}  Above this temperature threshold, vortices of elasticity $1/\alpha$ are stable.  In the point vortex limit of $\alpha\rightarrow\infty$ (complete rigidity), the transition moves to infinite temperature and does not occur.  Two-dimensional vortices are unstable at all temperatures other than zero in this model because they have no internal entropy and fail to minimize the free energy.  As the temperature of the system increases, the system prefers more elastic vortices.

I have solved exactly the nearly parallel vortex filament model in a mean-field theory and have shown an infinite order phase transition which is not present in the 2-D point vortex gas model.  This novel result indicates that vortex formation and stability are highly dependent on core structure (viscosity and core velocity) reflected in $\alpha$.  I have also shown that the point vortex model is at best accurate at very low energies where vortex elasticity and entropy are negligible.

Future work includes Monte Carlo simulations of the microcanonical ensemble to determine if the specific heat is negative at the transition point.  Lynden-Bell has shown that systems with long-distance interaction potentials may have negative specific heat \cite{Lynden:1968, Lynden:1977}.  Systems that experience negative specific heat in the microcanonical ensemble for a range of energies undergo a phase transition in the canonical ensemble for that range \cite{Thirring:1970}.  Preliminary results indicated that this system experiences negative specific heat, but, until now, we have not known the precise transition point \cite{Andersen:2007b}.  Given the canonical phase transition at $T_c$, the system will have negative specific heat when its mean temperature crosses this point in the fixed energy ensemble.  As in stellar systems, this point represents a meta-stable regime where the system raises its temperature by releasing energy and collapses into a more stable configuration.

\bibliography{pimcposter}
\end{document}